\documentclass[pre,twocolumn,superscriptaddress]{revtex4}
\usepackage{amsmath}
\usepackage{bm}
\usepackage{graphicx} 
\usepackage{color}

\begin{document}

\title{How active forces influence nonequilibrium glass transitions} 

\author{Ludovic Berthier}
\affiliation{Laboratoire Charles Coulomb, UMR 5221 CNRS, 
Universit{\'e} Montpellier, Montpellier, France}

\author{Elijah Flenner}
\affiliation{Department of Chemistry, Colorado 
State University, Fort Collins, CO 80523, USA}

\author{Grzegorz Szamel}
\affiliation{Laboratoire Charles Coulomb, UMR 5221 CNRS, 
Universit{\'e} Montpellier, Montpellier, France}

\affiliation{Department of Chemistry, 
Colorado State University, Fort Collins, CO 80523, USA}

\date{\today}


\begin{abstract}
Dense assemblies of self-propelled particles undergo a nonequilibrium form of glassy 
dynamics. Physical intuition suggests that increasing departure from equilibrium due to 
active forces fluidifies a glassy system. We falsify this belief by 
devising a model of self-propelled particles where increasing departure from 
equilibrium can both 
enhance or depress glassy dynamics, depending on the chosen state point. We analyze 
a number of static and dynamic observables and suggest that the location of the 
nonequilibrium glass transition is primarily controlled by the evolution of 
two-point static density correlations due to active forces. The dependence of
the density correlations on the active forces varies
non-trivially with the details of the system, and is difficult to predict theoretically. 
Our results emphasize the need to develop an accurate liquid state theory for 
nonequilibrium systems. 
\end{abstract} 

\maketitle

\section{Introduction}

Nonequilibrium glass transitions have recently emerged 
as a new type of dynamic arrest occurring in particle 
systems driven out of equilibrium by active forces~\cite{Henkes2011,Angelini2011}. 
The initial theoretical interpretation, based on the analysis 
of simple glass models driven by active forces~\cite{Berthier2013}, 
has been confirmed in several computer simulations of more realistic active matter 
models~\cite{Henkes2011,Ni2013,Berthier2014,Wysocki2014,Fily2014,Szamel2015,joel,Mandal2016,Flenner2016,Bi2016,pinaki2017,chinese}. A number of alternative theoretical approaches have 
now been proposed to describe this 
phenomenon~\cite{Szamel2016,Feng2017,Liluashvili2017}.  
Just as in equilibrium~\cite{Mari2009,Ikeda2012},
nonequilibrium glass transitions bear no connection to  
the jamming transition~\cite{jamming}, which corresponds instead 
to a geometric transition taking place in the absence of any driving
mechanism. 

Although the system is driven far from thermal equilibrium, 
the corresponding slow dynamics exhibits all the 
characteristic signatures of supercooled liquids 
approaching an equilibrium glass 
transition~\cite{Berthier2011}, such as caging, dynamical
slowing down, non-exponential time correlation functions 
and dynamic heterogeneity~\cite{Berthierbook}. A unique feature that 
distinguishes active from equilibrium glasses is the emergence
of collective effective temperatures~\cite{Berthier2013,Levis2015}.
Nonequilibrium glass transitions represent an experimentally relevant
concept, because they may explain various dynamic phenomena observed 
experimentally in both dense active colloidal suspensions~\cite{Ginot2015,clemens}, 
active granular materials~\cite{Deseigne2010}, and in biological 
systems~\cite{Angelini2011,Schotz2013,Garcia2015,Gravish2015,alberto}. 

The above summary demonstrates that the existence of nonequilibrium glassy dynamics is 
well-established.
For a number of model systems, phase diagrams, microstructure, dynamic 
timescales and length scales have been thoroughly analyzed. However, not much is known 
quantitatively about how active forces
influence the glass transition. In a very trivial sense, 
adding active forces to an equilibrium material must
suppress the glass transition, as the amount of driving
energy then increases~\cite{Henkes2011,Mandal2016}. This is equivalent to 
increasing the temperature at equilibrium. However, the outcome 
of departing from thermal equilibrium 
{\it at constant driving energy} is much less trivial, 
and is in fact not understood. There exist conflicting results in the literature, 
suggesting that 
glassy dynamics is either suppressed 
(as in hard sphere systems~\cite{Ni2013,Berthier2014}), or enhanced
(as in Lennard-Jones particles~\cite{Szamel2015,Flenner2016}) 
when going out of equilibrium.
It is unclear whether these seemingly distinct behaviors are due to a change in the pair 
interaction, to the details of the active forces
or microscopic dynamics, or to a genuine physical effect. 
If real, then, these results beg the question as to what physical 
quantity is the main indicator to reveal how active forces modify the location of the 
glass transition. 

In this work, we devise a simple molecular dynamics model for self-propelled 
particles where acceleration or slowing down of the dynamics 
can both be observed by changing active forces at constant density of the system.
This directly shows that it is in fact very difficult to predict whether active 
forces will fluidify or glassify a given material, as they can do both.  
Having a model displaying both types of response to active forces 
allows us to directly investigate how the departure from equilibrium influences the 
glassy dynamics, and which microscopic quantity is responsible for their evolution. Our 
numerical analysis reveals that active forces have a strong impact on 
the microstructure of the fluid, which can be readily quantified by two-point static 
density correlations functions. The nontrivial evolution of the static structure 
then primarily accounts for the evolution of the glass transition, underlying the 
need to develop a more accurate liquid state theory for active fluids.

The paper is organized as follows. In Sec. \ref{sec:model} we define our model system of 
self-propelled particles and provide the details of the numerical simulations. 
The main features of the glassy dynamics and the evolution of the apparent
glass transition line upon increasing departure from equilibrium are reported in 
Sec. \ref{sec:pd}. Next, in Secs. \ref{sec:velocity} and \ref{sec:density} 
we analyze two sets of equal-time steady-state 
correlations, velocity correlations and two-point density correlations, respectively. 
We conclude in Sec. \ref{sec:disc} with a discussion of the correlations between the 
evolution of the glassy dynamics and of the steady-state structure upon
increasing departure from equilibrium.

\section{Interpolating between hard and soft
active particles}\label{sec:model}

Earlier studies of glassy dynamics in model active systems used either a
hard-sphere interaction~\cite{Ni2013,Berthier2014} or a Lennard-Jones 
interaction \cite{Szamel2015,Mandal2016,Flenner2016}. Both families
of studies reported opposite results regarding the 
influence of active forces on the glass transition. 

To continuously interpolate between these limiting cases, we 
use the Weeks-Chandler-Andersen (WCA) potential \cite{WCA},
a strategy used before in equilibrium studies~\cite{gilles}.
This choice allows us to continuously 
move from simulating a hard sphere-like system at very low temperatures and 
moderate densities (when the typical nearest-neighbor distance is slightly larger
than the range of the potential), to simulating a Lennard-Jones-like system
at moderate temperatures and large densities (when the typical nearest-neighbor 
distance is smaller that the range of the potential). 
The comparison with earlier works suggest that our model should display both 
an acceleration or a slowing down of the glassy dynamics, depending on the 
density regime, allowing us to revisit and unify previous studies. 

To model an active liquid, we 
use the so-called active Ornstein-Uhlenbeck particles model \cite{Fodor2016}
introduced independently in Refs. \cite{Szamel2014} and \cite{Maggi2015}. 
In this model, the dynamics is overdamped and 
the particles move under the combined influence
of the interparticle interactions and the self-propulsion. 
The self-propulsion is modeled as an internal driving force evolving
according to the Ornstein-Uhlenbeck process. Thus, the equations of motion
are given by
\begin{eqnarray}
\label{eq:motion1}
\dot{\bm{r}}_i & = & \xi_0^{-1}\left[ \bm{F}_i + \bm{f}_i \right], \\
\label{eq:motion2}
\tau_p \dot{\bm{f}}_i & =  & \bm{f}_i + \bm{\eta}_i. 
\end{eqnarray}
In Eq.~\eqref{eq:motion1}, $\bm{r}_i$ is the position of particle $i$, $\xi_0$ 
is the friction coefficient of an isolated particle, $\bm{F}_i$ is the force 
acting on particle $i$ originating from the interactions, and 
$\bm{f}_i$ is the self-propulsion force acting on particle $i$. 
In Eq.~\eqref{eq:motion2}, $\tau_p$ is the persistence time of the self-propulsion and 
$\bm{\eta}_i$ 
is an internal Gaussian noise with zero mean and variance 
$\left<\bm{\eta}_i(t)\bm{\eta}_j(t^\prime)\right>_{\mathrm{noise}} = 
2 \xi_0 T_\text{eff} \bm{I} \delta_{ij}\delta(t-t^\prime)$,
where $\left<...\right>_{\mathrm{noise}}$ denotes averaging over the noise 
distribution, $T_\text{eff}$ is the single-particle effective temperature, 
and $\bm{I}$ is the unit tensor. 
In the following, we set the friction coefficient to unity, $\xi_0 = 1$. 
Notice that $\bm{f}_i$ is the unique driving force in Eq.~(\ref{eq:motion1}),
which does not contain an additional Brownian noise term. 

The name `single-particle effective temperature' for $T_{\rm eff}$ originates from the fact that
an isolated particle moving under the influence of the self-propulsion 
evolving according to Eq. (\ref{eq:motion2}) performs a persistent random walk
with the long-time diffusion coefficient equal to $D_0=T_\text{eff}$
(we use a system of units such that the Boltzmann constant $k_B$ is unity). We note that
for a system of interacting self-propelled particles other effective temperatures 
can be defined based on different fluctuation-dissipation ratios~\cite{Berthier2013,Levis2015,Szamel2017}. These effective temperatures are, in general, different 
from the single-particle effective temperature. Since we will not
be concerned with these collective effective temperatures in this work, in the 
following, for brevity, we will refer to $T_\text{eff}$ as the effective
temperature. Importantly, $T_\text{eff}$ controls the amount of energy injected into the system, 
and it represents the analog of the thermal bath for an equilibrium system. 

The interparticle forces originate from a potential, 
$\bm{F}_i = -\sum_{j \ne i} \nabla_i V_{\alpha \beta}(r_{ij})$ where
\begin{equation}
V_{\alpha \beta}(r) = 4 \epsilon 
\left[ \left( \frac{\sigma_{\alpha \beta}}{r} \right)^{12} 
- \left( \frac{\sigma_{\alpha \beta}}{r} \right)^6 \right],
\end{equation}
and $\alpha, \beta$ denote the particle species 
$A$ or $B$, $\epsilon =1$ (which sets the unit of energy), 
$\sigma_{AA} = 1.4$, $\sigma_{AB} = 1.2$, and
$\sigma_{BB} = 1.0$ (which sets the unit of length). 
We simulated $N = 1000$ particles composing a 
50:50 mixture in a volume $V$ using periodic boundary conditions in three spatial dimensions. 
Following the WCA prescription, the potential is truncated and shifted at the
minimum, \textit{i.e.} at 
$\varsigma_{\alpha \beta} = 2^{1/6} \sigma_{\alpha \beta}$. Thus, the 
interparticle force is purely repulsive. The repulsive character of the force combined 
with the finite range of the potential implies that in the low temperature limit
the system becomes equivalent to a hard sphere system consisting of 
a binary mixture of spheres of
diameters $\varsigma_{AA}$ and $\varsigma_{BB}$.

As our control parameters, we use the volume fraction 
$\phi = \pi N [\varsigma_{AA}^3 + \varsigma_{BB}^3]/(12 V)$, the effective temperature 
$T_\text{eff}$, and the persistence time of the self-propulsion $\tau_p$. 
Since there is no thermal noise, when $\tau_p \rightarrow 0 $ this model system 
becomes equivalent to a Brownian system at a temperature $T = T_\text{eff}$. Therefore, $\tau_p$ quantifies the 
increasing departure from equilibrium as $\tau_p$ increases from zero. In this work, we investigate the dependence
of the glassy dynamics on the persistence time and we also compare the results obtained for active systems with 
those obtained from overdamped Brownian dynamics (BD) simulations at a temperature $T$. We vary the persistence 
time between $\tau_p=0$ and $\tau_p=10$ and the effective temperature between 
$T_\text{eff}=0.01$ and $T_\text{eff}=1.0$.

\section{Glassy dynamics and phase diagram}\label{sec:pd}

\begin{figure}
\includegraphics[width=8.5cm]{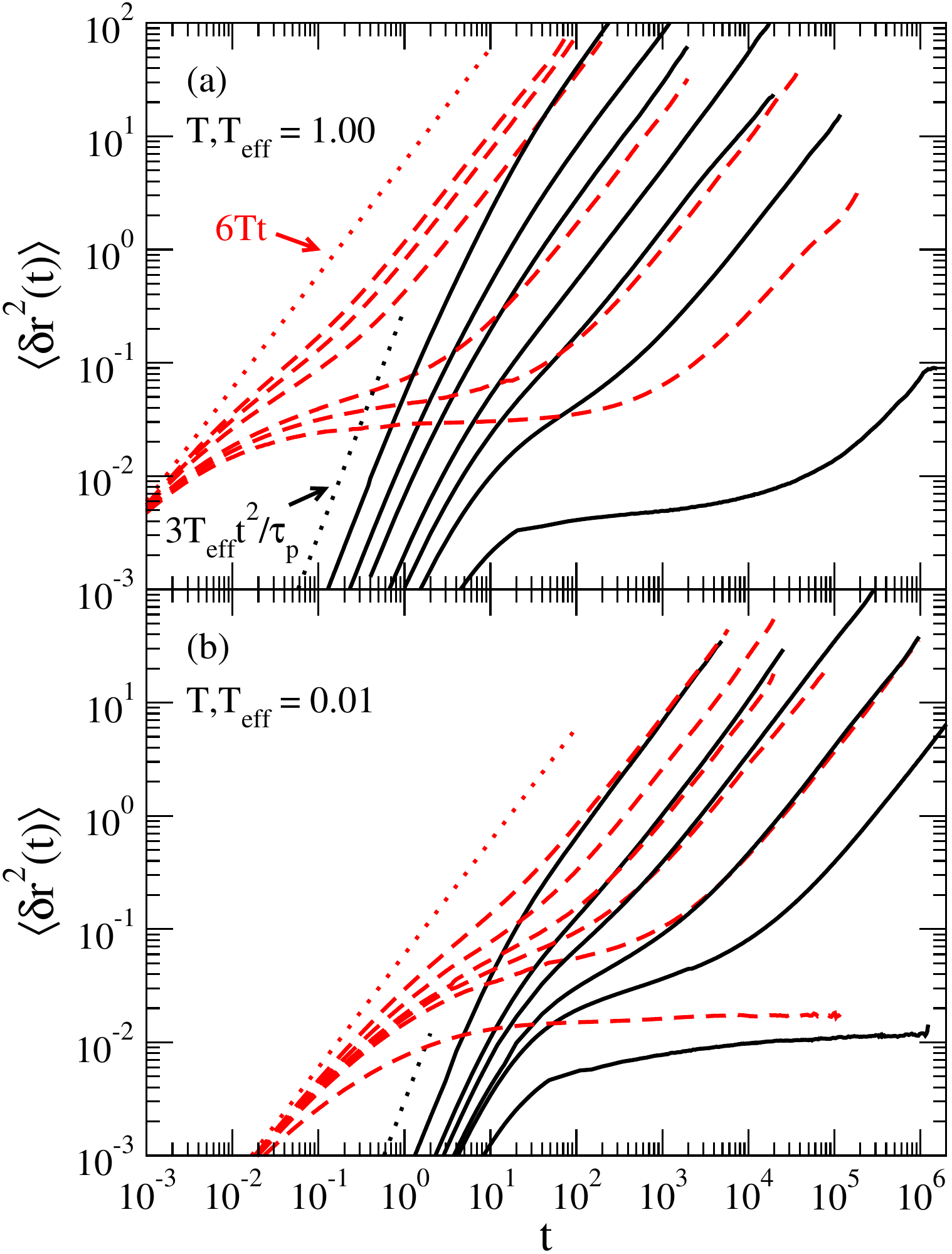}
\caption{\label{fig:msd} Time-dependence of the mean-squared displacement
for a high (a) and low (b) temperature. Black solid lines represent active systems for persistence time $\tau_p=10$ 
and red dashed lines represent Brownian systems ($\tau_p=0$). Dotted lines represent short-time 
motion of an isolated particle for active (black) and Brownian 
(red) systems.
In panel (a), solid lines represent active systems at
$\phi=0.554$, 0.638, 0.665, 0.693, 0.707, 0.721, 0.873 (left to right) and dashed lines represent Brownian systems at 
$\phi=0.638$, 0.693, 0.762, 0.873, 0.901, 0.928 (left to right). In panel (b), solid lines represent active 
systems at $\phi=0.554$, 0.624, 0.638, 0.652, 0.658, 0.665 (left to right)
and dashed lines represent Brownian systems at $\phi=0.499$, 0.554, 0.582, 0.596, 0.610, 0.652 (left to right).}
\end{figure}

In Fig.~\ref{fig:msd} we show the development of glassy dynamics upon increasing
the volume fraction, at two values of the effective temperature, 
the highest and the lowest temperature 
investigated, $T_\text{eff}=1.0$ and $T_\text{eff}=0.01$, respectively. 
We illustrate the changes in the dynamics 
by showing the mean-squared displacement 
\begin{equation}
\left< \delta r^2(t) \right> = \frac{1}{N_A} 
\left< \sum_{i=1}^{N_A} \left[\bm{r}_i(t) - \bm{r}_i(0)\right]^2 \right>,
\end{equation}
where the summation is over the particles of type $A$ and $N_A$ is the number of these 
particles. 
Here and in the following, 
we restrict our discussion to the larger particles, particles $A$, 
and we note that analyzing 
the dynamics of the $B$ particles leave the conclusions unchanged. 
Furthermore, to simplify notation we do not use conventional subscripts when
referring to quantities pertaining to particles $A$ only. Thus, we use
$\left< \delta r^2(t) \right>$ rather than $\left< \delta r^2_A(t) \right>$ and
later in Sec.~\ref{sec:density} we will use, for instance, $g(r)$ rather than $g_{AA}(r)$. 

For an isolated active particle or, alternatively, in a non-interacting system, 
the mean-squared displacement can be calculated analytically,
\begin{equation}
\left< \delta r^2(t) \right> = 6 T_\text{eff} 
\left[ \tau_p \left( e^{-t/\tau_p} - 1\right) + t \right].
\end{equation}
At short times, the particle motion is ballistic 
and $\left< \delta r^2(t) \right> \simeq 3 (T_\text{eff}/\tau_p)  t^2$. 
The long-time motion is diffusive and 
$\left< \delta r^2(t) \right> \simeq 6 T_\text{eff} t$. Comparing the long-time
result with that for an isolated Brownian particle, 
$\left< \delta r^2(t) \right> = 6 T t$, we see that the long time diffusive motion of 
an isolated active particle at an effective temperature $T_\text{eff}$ matches that of the 
isolated Brownian particle at a temperature $T=T_\text{eff}$. 

As shown in Fig.~\ref{fig:msd}, the ballistic and diffusive regimes are 
still observed in mean-squared displacements
in systems of interacting active particles (solid lines, $\tau_p=10$). 
However, for active particles both the short-time dynamics and the 
long-time dynamics change with the volume fraction. The change in the short-time
dynamics is induced by correlations between active particles velocities and
positions, discussed further in Sec.~\ref{sec:velocity}. These correlations are
an important feature of active systems~\cite{Szamel2015}. At constant effective temperature, 
their magnitude decreases with decreasing persistence time and the
correlations vanish in the 
Brownian limit. Generally, for active systems 
both the short-time ballistic motion and the long-time
diffusive motion slow down with increasing volume fraction. The slowing down of  the
short-time dynamics is more pronounced at higher effective temperatures (note, however,
that at higher effective temperatures the volume fractions are also somewhat larger). 
Generally, with increasing volume fraction, at intermediate times a plateau begins to 
develop and glassy dynamics emerge.

Figure~\ref{fig:msd} also shows that for Brownian systems we have two diffusive regimes, 
for short times $\left< \delta r^2 (t) \right> =   6 T t$ (dotted red line)
and for long times $\left< \delta r^2(t) \right> = 6 D t$, where $D$ is the long-time
self-diffusion coefficient. At intermediate times a
plateau develops between the short time diffusive motion and the long time
diffusive motion. The presence of the plateau indicates caging of individual particles. 

While emerging glassy dynamics in active systems is generally similar to that in
Brownian systems, we note some important quantitative differences. First, in active
systems there is a significant slowing down of the short-time ballistic motion
whereas in Brownian systems the short-time diffusive motion is independent of the
volume fraction (and it depends only trivially on the temperature). Second,
plateau heights in active systems are different from those in Brownian systems. 
This is especially prominent at the higher temperature where the plateau height
for the densest $\tau_p = 10$ active system is around an order of
magnitude smaller than for the densest Brownian system (note that the density
of the active system is quite a bit lower than that of the Brownian system).
For $T_{\rm eff}=0.01$, a well-defined plateau is not observed and instead the mean-squared
displacement exhibit a very slow subdiffusive behavior.
These latter facts suggest that upon increasing departure from equilibrium the 
effective interparticle interaction changes significantly. 
We comment on this point further in Sec.~\ref{sec:density}. 

\begin{figure} \hspace*{-0.3cm}
\includegraphics[width=8.2cm]{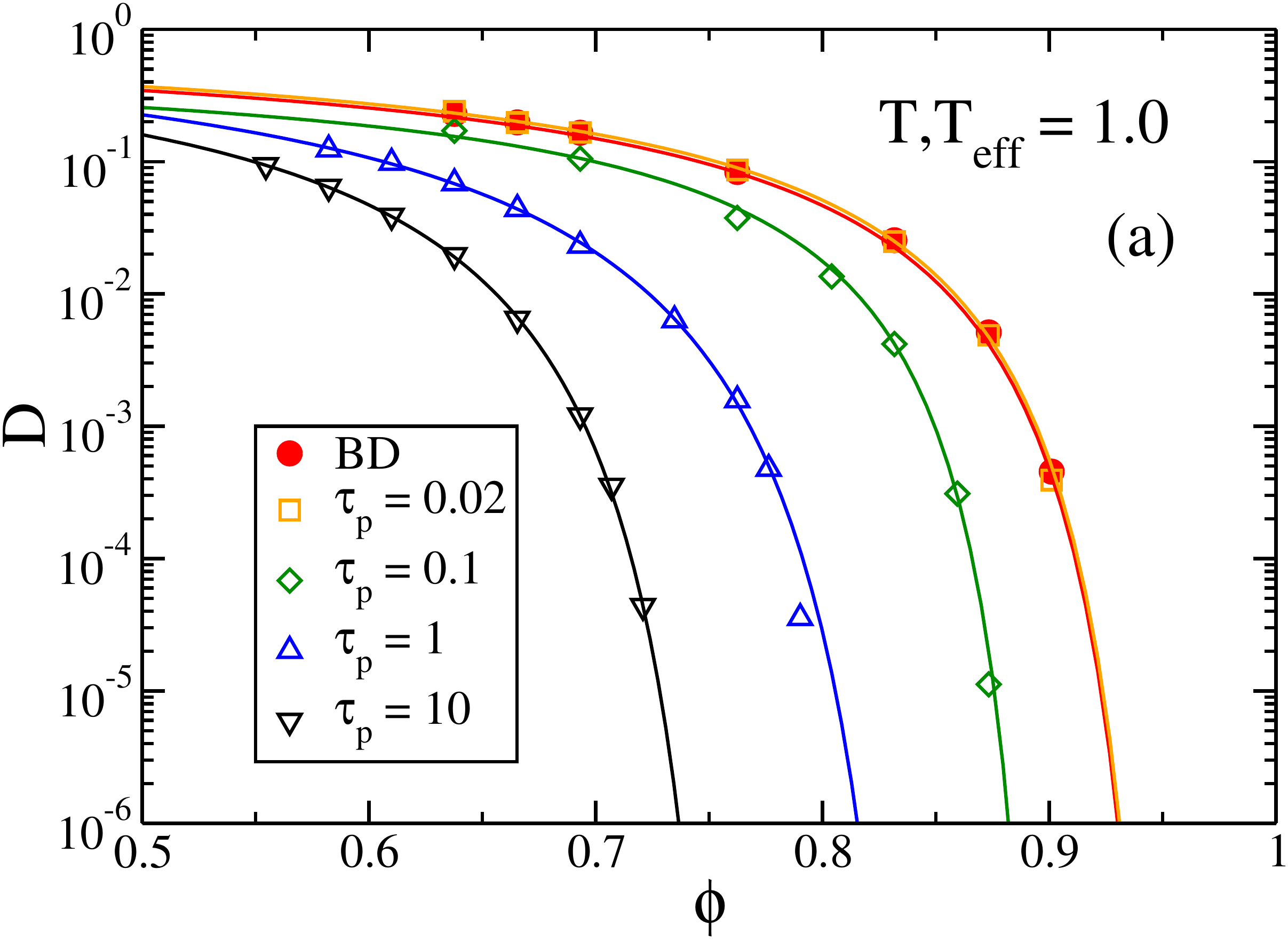}
\includegraphics[width=8.5cm]{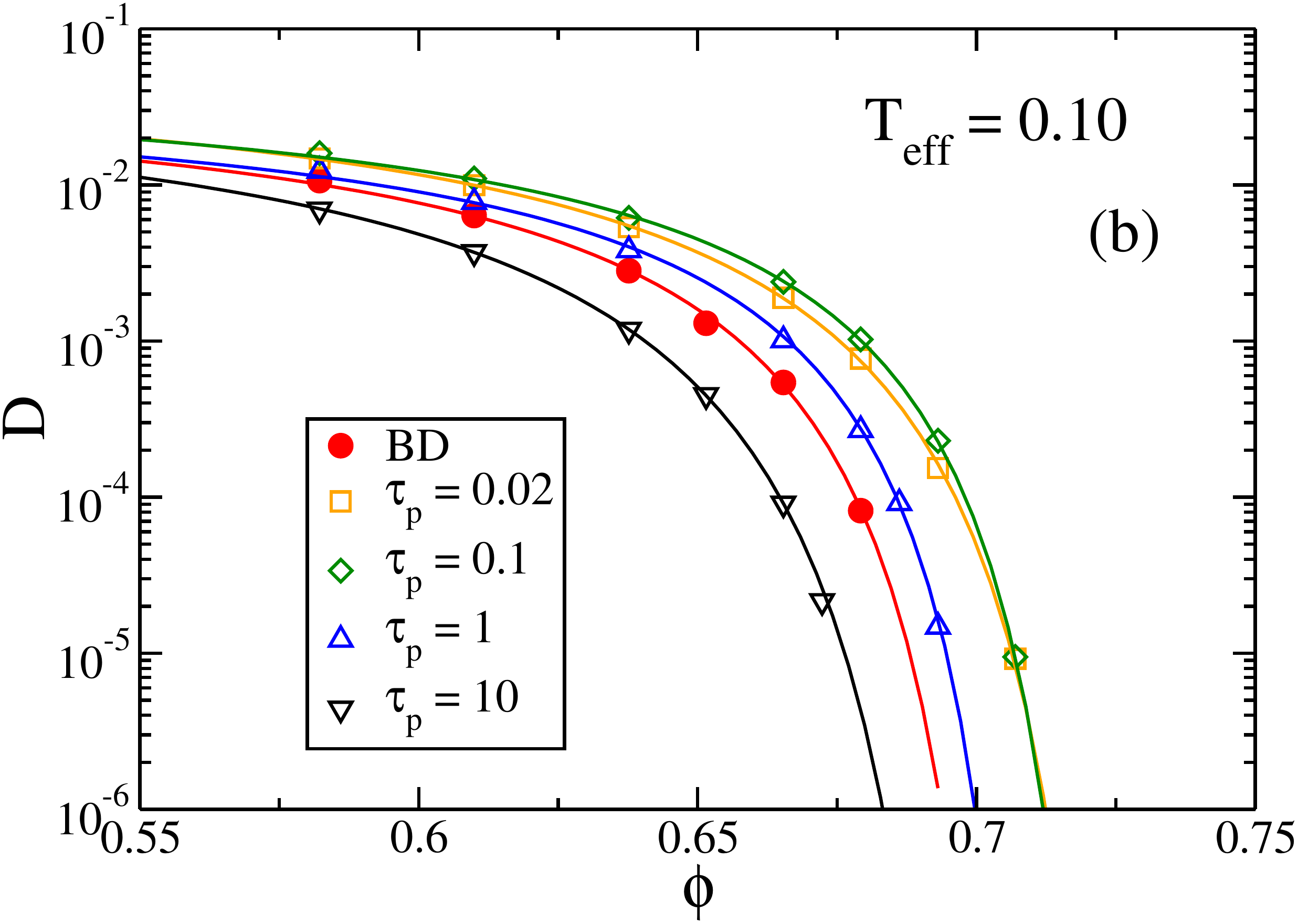}
\includegraphics[width=8.5cm]{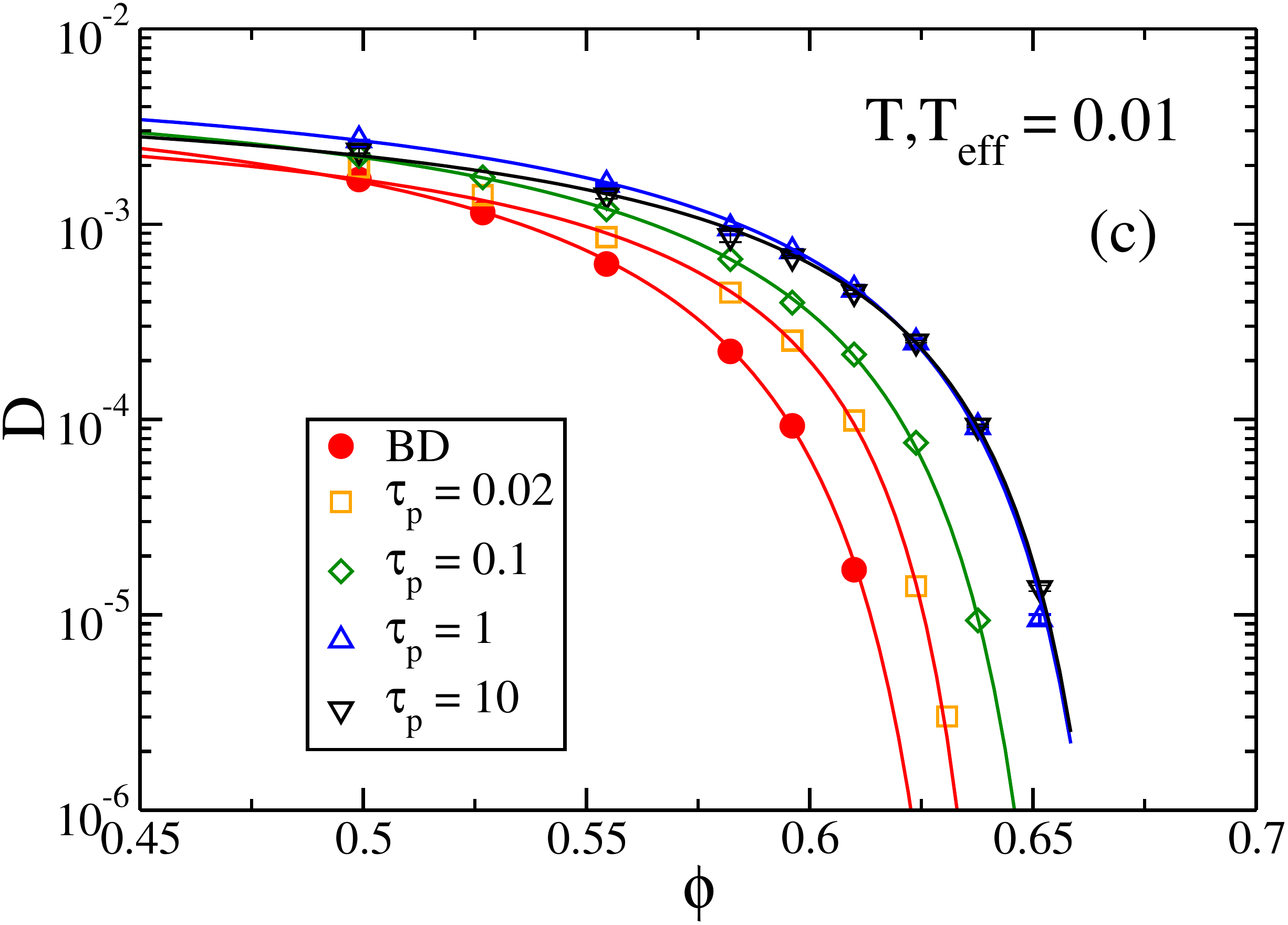}
\caption{\label{fig:diffusion} 
Volume fraction dependence of the long-time self-diffusion coefficient at
different persistence times at three representative temperatures.
The points represent results of numerical simulations and the lines are 
Vogel-Fulcher-like fits, $\ln D = A+B/(\phi-\phi_c)$.
For each system glassy dynamics is observed 
as the volume fraction increases. The glass transition is shifted 
to larger $\phi$ as the persistence time increases at low
$T_\text{eff}$ (a), decreases at large $T_\text{eff}$ (c), and is non-monotonic
at intermediate $T_\text{eff}$ (b). Filled symbols are Brownian dynamics simulations.}
\end{figure}

From the long-time limit of the mean-squared displacements 
we extract the long-time self-diffusion coefficients, 
$D=\lim_{t\to\infty} \left< \delta r^2 (t) \right> / (6t)$. 
In Fig.~\ref{fig:diffusion}, we show the dependence of the diffusion 
coefficients on the volume fraction for a number of active systems
characterized by a given value of the persistence time and the effective temperature,
$(\tau_p, T_\text{eff})$, and for Brownian systems characterized by the temperature $T$. 
For the highest temperature. $T_\text{eff} = 1.0$, 
in Fig.~\ref{fig:diffusion}(a) we see that the diffusion coefficient 
decreases with increasing persistence time at a fixed volume fraction. In contrast,
at an intermediate temperature, $T_\text{eff} = 0.1$, we find a non-monotonic 
dependence of the diffusion coefficient on the persistence time at a fixed volume fraction. 
Finally, at the lowest temperature, $T_\text{eff} = 0.01$, we find that the diffusion 
coefficient increases with increasing persistence time at a fixed volume fraction for 
the range of persistence times investigated. We observe that increasing departure 
from equilibrium can either promote or suppress the glassy dynamics without changing the 
pair interaction between the self-propelled particles, as announced in the introduction.  
 
These results suggest that there is a change in the persistence time dependence of the 
apparent glass transition line in the temperature-volume fraction plane. 
To determine quantitatively the glass transition line, we 
fit the diffusion coefficient data to a Vogel-Fulcher-like dependence on the
volume fraction, $\ln D = A+B/(\phi-\phi_c)$, where $A$, $B$ and $\phi_c$ are fitting parameters. 
These empirical fits are shown as continuous lines in Fig.~\ref{fig:diffusion}. 
The Vogel-Fulcher-like formula results in reasonable fits 
to the data and, therefore, reasonable estimates for the glass transition volume 
fraction $\phi_c=\phi_c(\tau_p,T_{\rm eff})$. Other fitting functions would provide 
qualitatively similar results for the evolution of the glass transition lines.
 
\begin{figure}
\includegraphics[width=8.5cm]{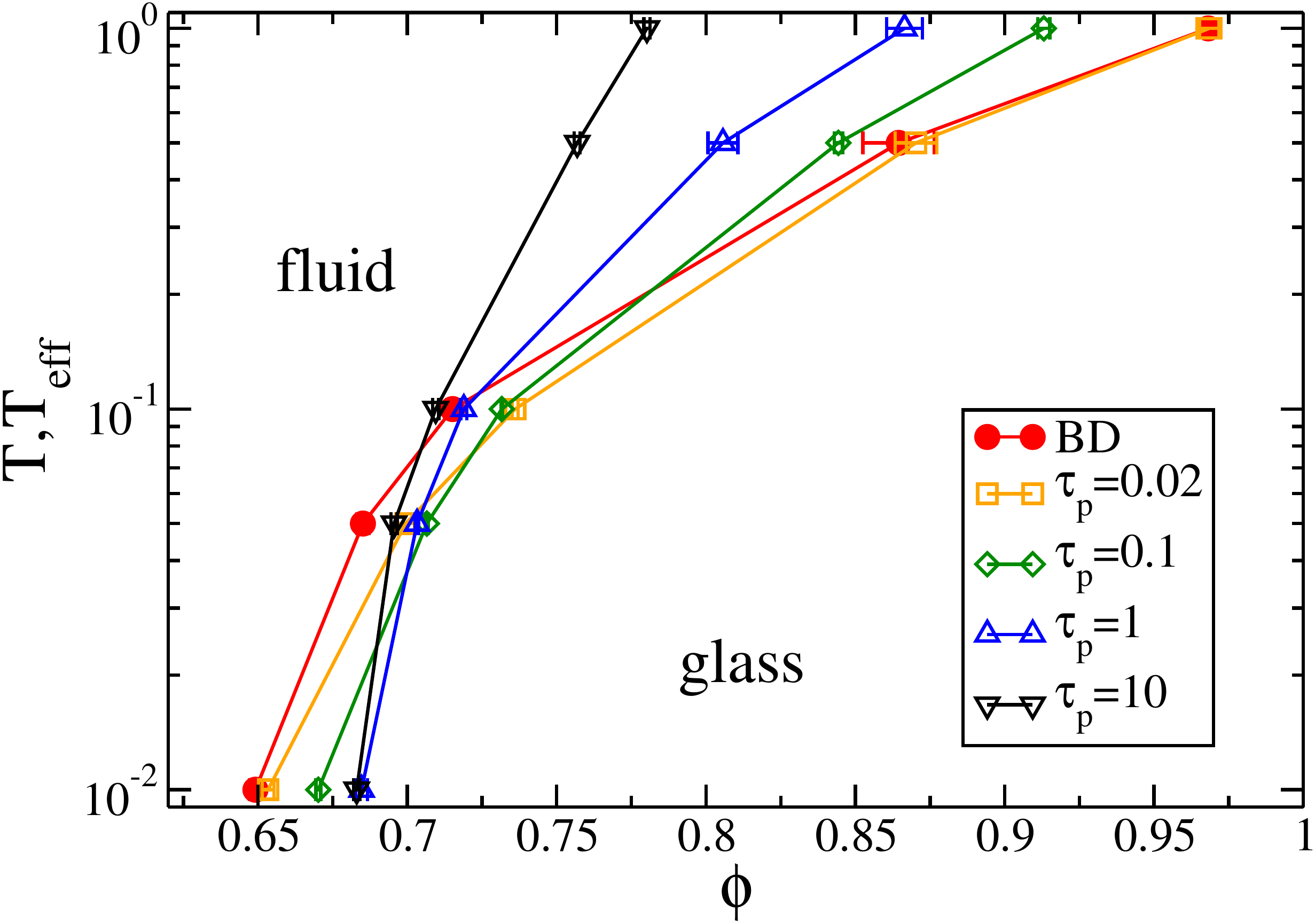}
\caption{\label{fig:phasediagram}
Evolution of the phase diagram with the persistence time of the self-propulsion. 
The fluid phase at low $\phi$ transforms into a glass
at large $\phi$ at a volume fraction which depends both 
$T_\text{eff}$ and $\tau_p$. With increasing persistence time the glass transition
lime shifts towards smaller volume fractions at higher effective temperatures 
(departure from equilibrium promotes glassy dynamics) and towards larger volume 
fractions at low effective temperatures (departure from equilibrium fluidifies the 
glass). Filled symbols are Brownian dynamics simulations.}
\end{figure}

In Fig.~\ref{fig:phasediagram} we present the resulting glass-fluid 
phase diagram in the temperature-volume 
fraction plane, for different persistence 
times. As can be inferred from the dependence of the diffusion coefficient on the 
volume fraction, the glass transition volume fraction $\phi_c$ monotonically 
decreases with increasing $\tau_p$ for $T_\text{eff} = 1.0$, and it monotonically 
increases with increasing $\tau_p$ for $T_\text{eff} = 0.01$. 
At intermediate $T_\text{eff}$ there is a non-monotonic change of $\phi_c$ upon
increasing the persistence time, which signals the crossover between the high and low temperature regimes. 
These findings are consistent with both the previously observed increase of the glass transition temperature 
for the Lennard-Jones system~\cite{Flenner2016}
and the increase of the glass transition packing fraction for the hard-sphere system~\cite{Ni2013,Berthier2014}. 
 
Figure~\ref{fig:phasediagram} shows for the first time that for the same model active system increasing 
departure from equilibrium can both glassify an active fluid and fluidifies an active glass, depending on 
the studied thermodynamic state point. This directly demonstrates that the physical intuition that activity 
fluidifies the glass is incorrect, as activity can also solidify the supercooled fluid.  
In the following two sections we search for correlations
between the observed non-trivial dependence of the dynamics on the persistence time and the
dependence of static (equal-time) correlations on the persistence time. 
In Sec.~\ref{sec:velocity} we examine the evolution of the correlations between 
active particles' velocities on the persistence time. In Sec.~\ref{sec:density}
we investigate the dependence of density fluctuations on the persistence time.  

\section{Nonequilibrium velocity correlations}
\label{sec:velocity}

While developing a theory for the dynamics of systems of active Ornstein-Uhlenbeck
particles \cite{Szamel2015,Szamel2016} we discovered that correlations
between velocities of self-propelled particles play an important role in the 
dynamics of active systems. Subsequent simulational studies \cite{Szamel2015,Flenner2016}
showed that in the active system with Lennard-Jones interactions 
the velocity correlations grow upon increasing departure from equilibrium.
The correlations between velocities of difference active particles 
vanish in the limit of zero persistence time and do not exist in Brownian systems. 
Therefore, these nonequilibrium velocity correlations represent the most natural candidate to explain 
how the glass transition departs from its equilibrium counterpart as the persistence time increases.

In Refs. \cite{Szamel2015,Szamel2016}, we introduced   
a wavevector-dependent function characterizing velocity correlations and showed that 
this function is important 
for the dynamics of an active system. For a binary mixture, in analogy with the
partial static structure factors that characterize the number density fluctuations, 
in principle we need to introduce
three different functions corresponding to the correlations of velocities of
the $A$ particles, velocities of the $B$ particles, and the 
$AB$ cross-correlations.
As stated earlier, we restrict the discussion to the larger 
particles and  only examine the correlation between velocities of
the $A$ particles, which is defined as follows:
\begin{equation}
\omega_{||}(q) = \hat{\bm{q}} \cdot 
\left< \sum_{i,j=1}^{N_A}\left(\bm{f}_i+\bm{F}_i\right)\left(\bm{f}_j+\bm{F}_j\right)
e^{-i\bm{q}(\bm{r}_i-\bm{r}_j)}\right>\cdot\hat{\bm{q}},
\end{equation}
where both summations are over the particles of type $A$, 
$\hat{\bm{q}} = \bm{q} / | \bm{q} |$ and $\xi_0^{-1}\left(\bm{f}_i+\bm{F}_i\right)$ 
is the instantaneous velocity of particle $i$.

\begin{figure}
\includegraphics[width=8.5cm]{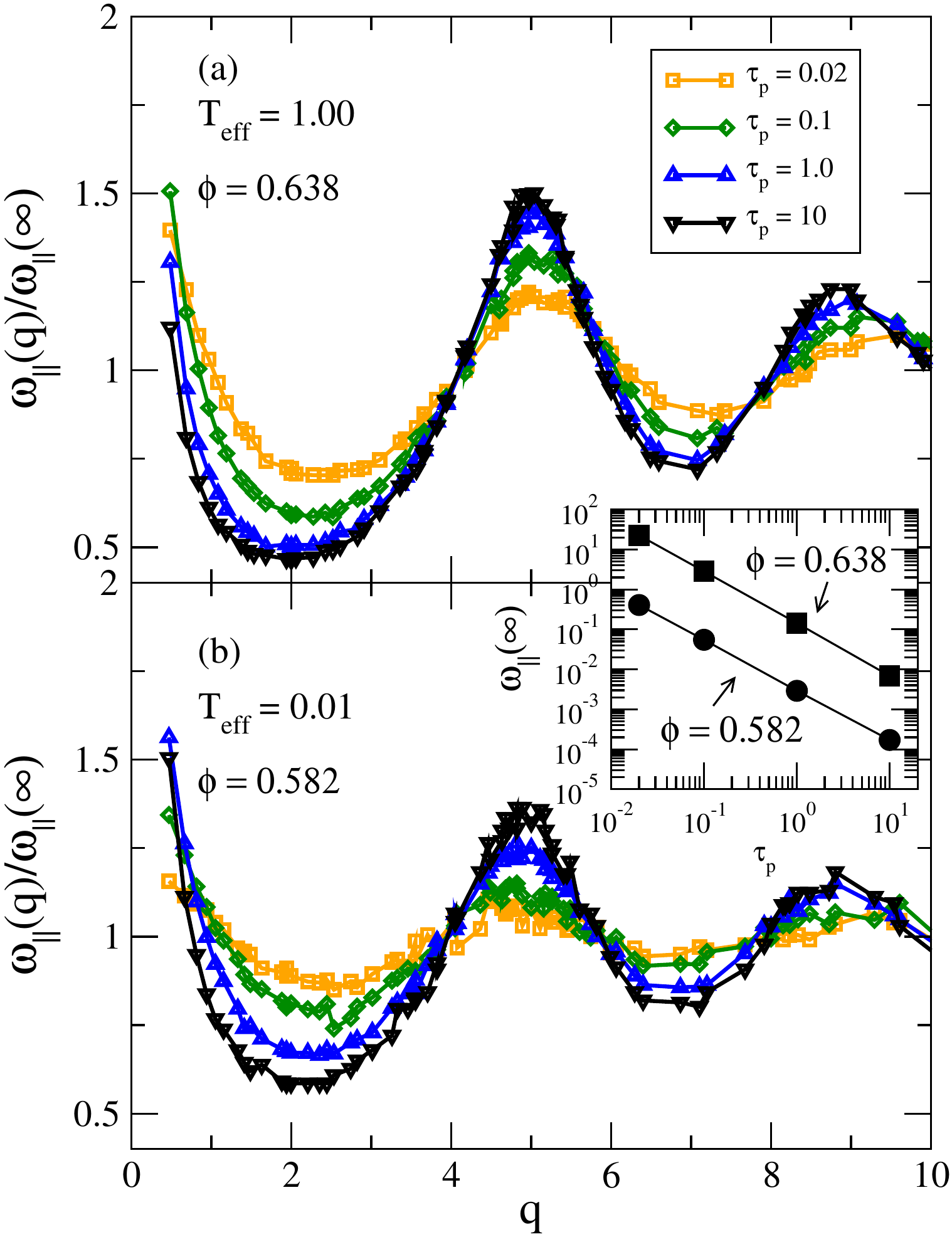}
\caption{\label{fig:velocity}
Wavevector dependence of the normalized nonequilibrium velocity correlations. These correlations develop and 
increase in strength 
as the persistence time increases, for both large and low $T_\text{eff}$, it is unity for Brownian dynamics. 
The inset show the persistence time dependence of the of $\omega_{||}(q \to \infty)$. Unlike the long-time glassy 
dynamics, the qualitative evolution of nonequilibrium velocity correlations does not depend on the state point.} 
\end{figure}

The large wavevector limit of this function, 
$\omega_{||}(\infty) = \lim_{q\to\infty} \omega_{||}(q)$, exactly determines the short-time 
behavior of the mean-squared displacement in the 
system of interacting active particles, 
$\left< \delta r^2(t) \right> \simeq 3 \omega_{||}(\infty)  t^2$.
According to the approximate mode-coupling-like theory sketched in Ref.~\cite{Szamel2015}
and then detailed in Ref. \cite{Szamel2016}, the complete function
$\omega_{||}(q)$ enters into the expressions for the so-called vertices of 
the irreducible memory function, and thus, together with the static structure factor, 
it determines the long-time dynamics. If the function $\omega_{||}(q)$ evolved differently  
upon increasing the persistence time at high and low effective temperatures (\textit{i.e.}
for Lennard-Jones-like and hard-sphere-like systems), it would suggest that the velocity 
correlations are responsible for the non-trivial changes in the phase diagram shown in 
Fig.~\ref{fig:phasediagram}.

Previous work \cite{Flenner2016} showed that upon increasing both the 
persistence time and the importance of the interparticle interactions, 
the overall scale of the velocity correlations decreases significantly
while their wavevector dependence becomes more pronounced. The overall scale of the
velocity correlations can be characterized by their large wavevector limit,
$\omega_{||}(\infty)$. We recall that Fig.~\ref{fig:msd} indicates that 
the short-time dynamics of the self-propelled particles slows down upon
increasing the volume fraction at fixed persistence time, for both
high and low effective temperatures.
This slowing down of the short-time dynamics is the direct consequence of the 
decreasing of $\omega_{||}(\infty)$ upon increasing the volume fraction at 
constant persistence time and effective temperature, and   
thus increasing the importance of the interparticle interactions.

The focus of this work is the dependence of the 
(long-time) glassy dynamics on the 
departure from equilibrium. To this end, 
in the inset to Fig.~\ref{fig:velocity} we show the evolution of 
$\omega_{||}(\infty)$ with the persistence time for $T_\text{eff} = 1.0$ and 0.01. 
In both cases we observe approximate power-law dependence of $\omega_{||}(\infty)$
on $\tau_p$, $\omega_{||}(\infty)\propto \tau_p^{-1.25}$. Thus, while 
$\omega_{||}(\infty)$ at $T_\text{eff} = 1.0$ is about 50 times larger
than that at $T_\text{eff} = 0.01$, its dependence on the persistence time
is the same at both temperatures, 
and, therefore, $\omega_{||}(\infty)$ seems unconnected to the more complicated 
evolution of the glassy dynamics reported in Figs.~\ref{fig:diffusion}(a-c).

In the main panels of Fig.~\ref{fig:velocity} we show the evolution of the 
wavevector dependence of $\omega_{||}(q)$ upon increasing the persistence time 
for $T_\text{eff} = 1.0$ and 0.01. 
To this end we plot $\omega_{||}(q)/\omega_{||}(\infty)$ for different values
of $\tau_p$. Upon increasing the persistence time 
we observe that the oscillations of 
$\omega_{||}(q)/\omega_{||}(\infty)$ are becoming more pronounced, which  indicates 
growing local velocity correlations. However, this increase occurs at every effective 
temperature investigated, independently of the evolution of the long-time glassy dynamics.  
Thus, growing local velocity correlations and glassy dynamics appear to be largely uncorrelated.
In other words, velocity correlations accompany the nonequilibrium glass transition, they presumably 
quantitatively affect its location, but they do not seem to be the main factor responsible for the 
nontrivial evolution of the phase diagram shown in Fig.~\ref{fig:phasediagram}.

\section{Two-point density correlations}
\label{sec:density}

Since the dependence of the velocity correlations on the persistence time
is uncorrelated to the evolution of the glassy dynamics, we turn our
attention to other, more conventional static correlations, namely
density correlations. 
In this section we investigate equal-time correlations of the density fluctuations by examining 
the pair correlation function in real and Fourier space. We show that the dependence of pair 
correlations upon increasing the departure from equilibrium strongly correlates with the evolution 
of the glassy dynamics reported in Figs.~\ref{fig:diffusion}(a-c), and we provide a physical interpreation of the results. 
 
\subsection{Pair correlation function $g(r)$}

\label{sec:gr}

\begin{figure}
\includegraphics[width=8.5cm]{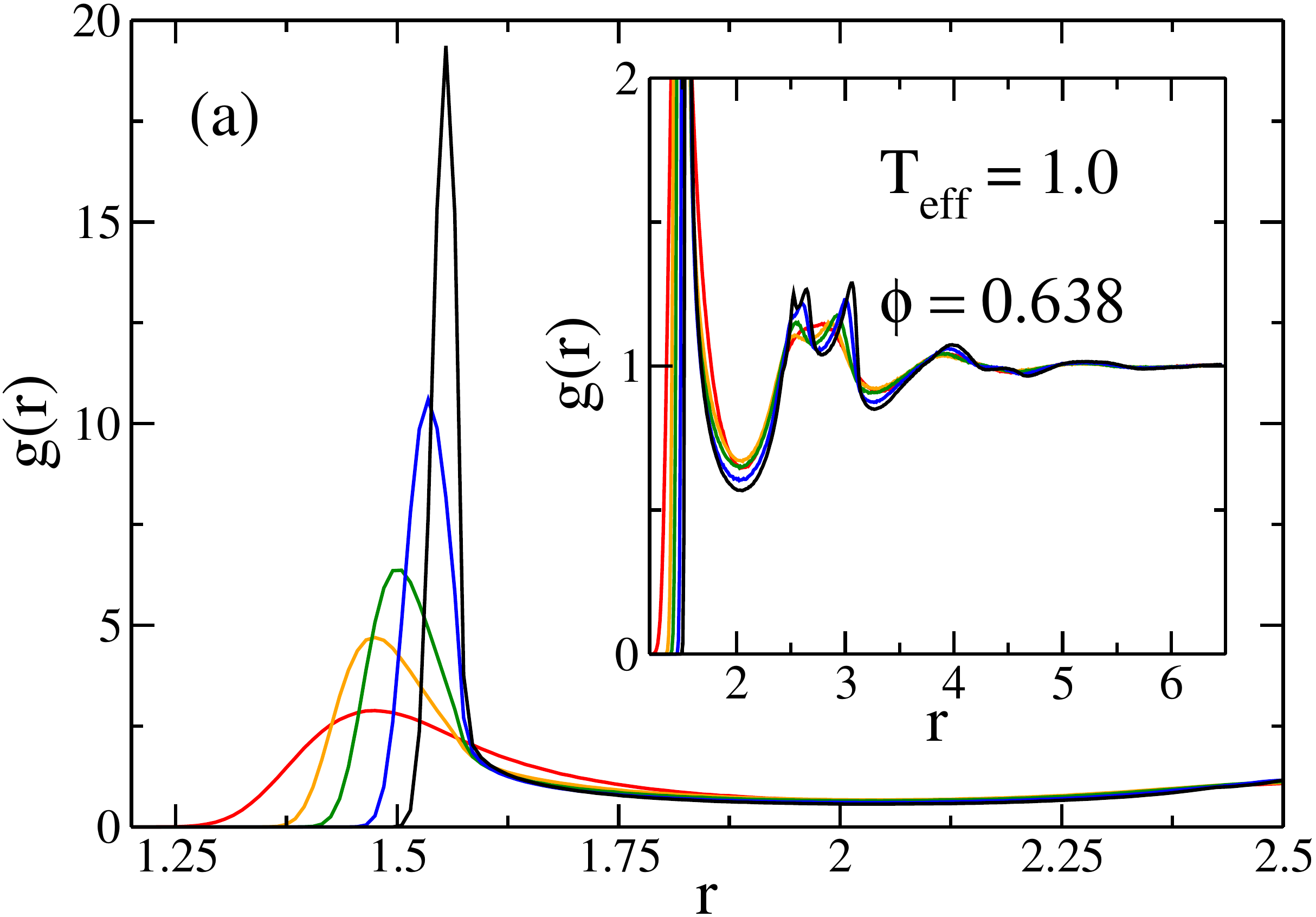}
\includegraphics[width=8.5cm]{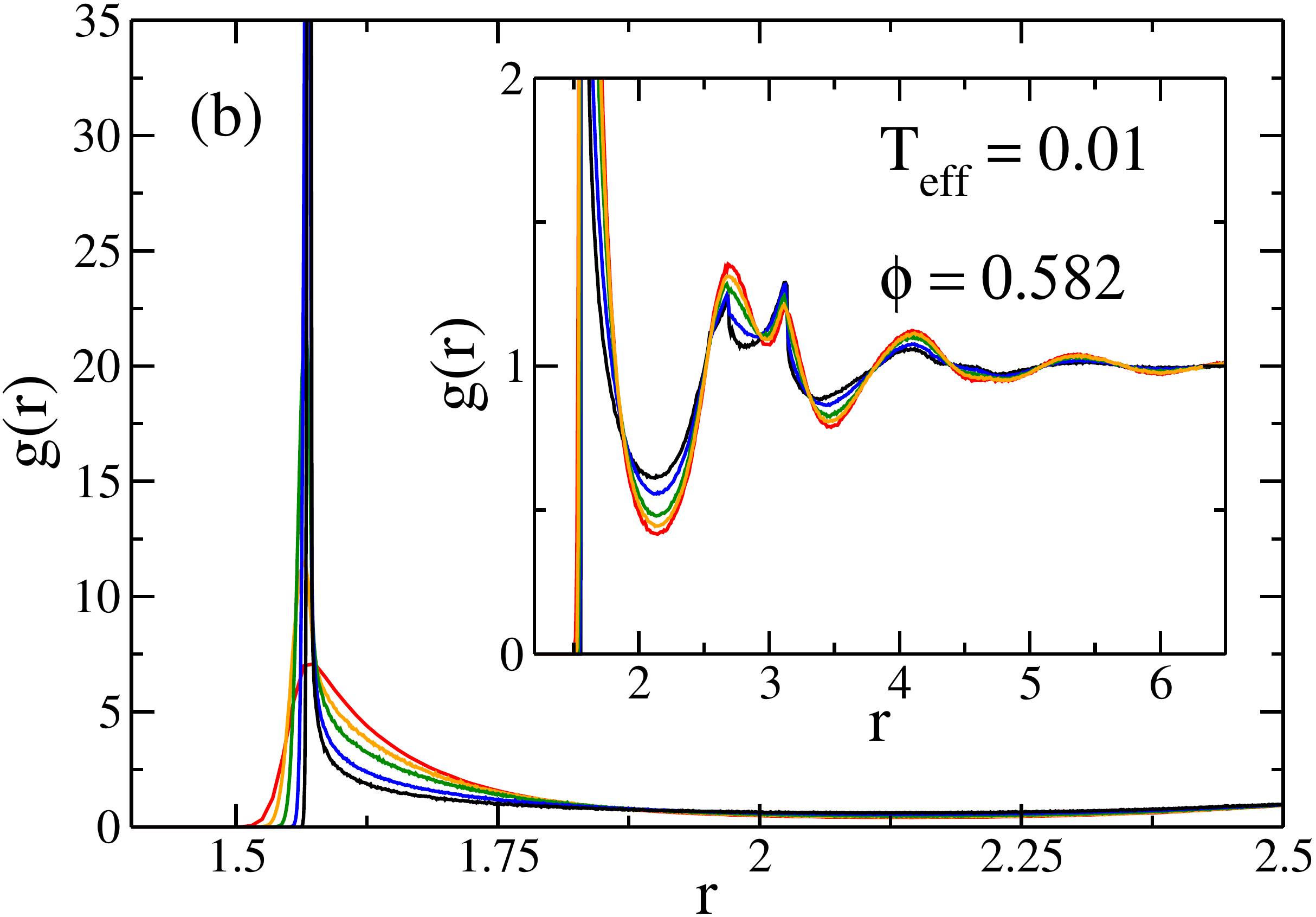}
\caption{\label{fig:gr} 
Evolution of the pair correlation function $g(r)$ with 
persistence time at $(T_\text{eff}=1.0, \phi=0.638)$ (a) and  
$(T_\text{eff}=0.01, \phi=0.582)$ (b),
using the same color coding for $\tau_p$ as in previous figures. 
The main panels show the first peak and the insets focus on the secondary peaks at larger $r$.}
\end{figure}

In Fig.~\ref{fig:gr} we show the evolution of pair density correlations in the
direct space upon increasing the persistence time. Specifically, we show
the pair correlation function characterizing correlations between the $A$ particles, 
\begin{equation}
g(r) = \frac{1}{N_A} \sum_{i=1}^{N_A} \sum_{j\ne i}^{N_A} 
\left< \delta(r-|\bm{r}_i-\bm{r}_j|)\right>,
\end{equation} 
where both summations are over the $A$ particles. 

For both high and low effective temperatures we observe a 
very strong sharpening and growth of the
first peak of the pair correlation function. This may be interpreted as an increasing 
`adhesion' between self-propelled particles upon increasing the persistence time.  The 
height of the first peak of $g(r)$ is 4.5 to 7 times larger for $\tau_p = 10$ than for 
the Brownian limit at the same packing 
fraction. While the quantitative details of the evolution of the first peak change with 
the effective temperature, qualitatively the evolution is very similar. Thus, the 
sharpening and growth of the first peak of the pair correlation function is a general 
feature of self-propelled particles which 
does not seem to be correlated with the non-trivial evolution of the
glass transition line. 

On the other hand, the position of the first peak has a very different behavior in the two 
panels shown in Fig.~\ref{fig:gr}. It shifts to larger distances for $T_{\rm eff} = 1.0$ 
but remains essentially at the same position for $T_{\rm eff}=0.01$. As a result, the 
effective diameter of the particle increases with $\tau_p$ for $T_{\rm eff} = 1.0$ but 
is approximately constant for $T_{\rm eff}=0.01$. This effective inflation of the particles 
for the higher temperature may be held responsible for the slowing down of the dynamics as 
$\tau_p$ increases, as the system becomes effectively more crowded. Strikingly, this effect is absent for the lower temperature. 

In the insets in Fig.~\ref{fig:gr}, we show that the secondary
peaks of the pair correlation function change in a qualitatively different  
way upon increasing departure from equilibrium. Specifically, the 
amplitude of the secondary peaks increases with
increasing $\tau_p$ for $T_\text{eff} = 1.0$ (indicating enhanced local structure) while it 
decreases with increasing $\tau_p$ for $T_\text{eff} = 0.01$ (indicating decreasing local 
structure). This qualitatively different change does correlate with the evolution of the
glass transition lines, where enhanced (suppressed) structure seems to promote (suppress) 
the glassy dynamics. As shown in the next subsection, the rather
subtle changes of the secondary peaks in the direct space translates into 
more visible changes of the static structure factor in the Fourier domain. 

\subsection{Static structure factor $S(q)$}

\begin{figure}
\includegraphics[width=8.5cm]{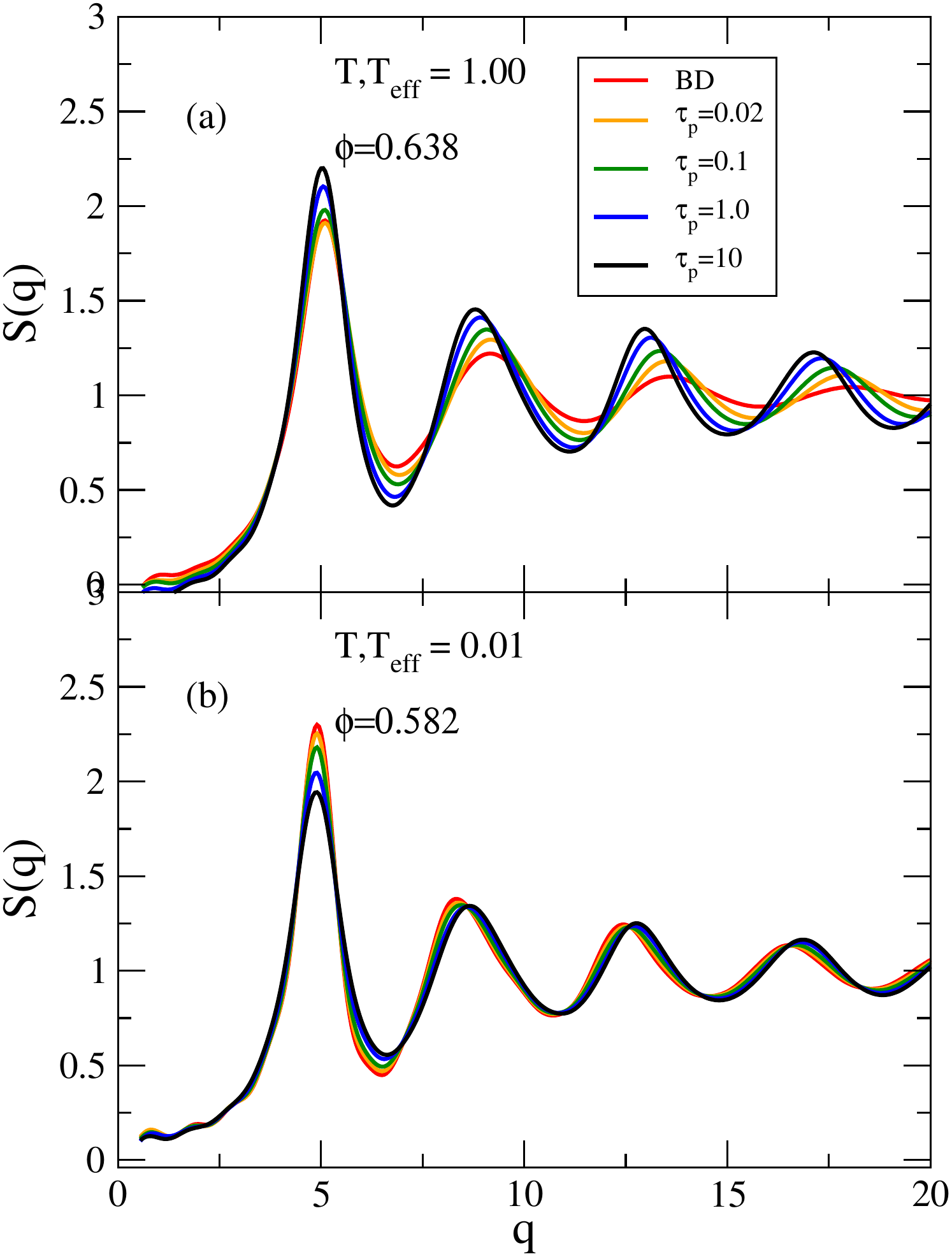}
\caption{\label{fig:sq}
Evolution of the static structure factor with the  
persistence time  at $(T_\text{eff}=1.0, \phi=0.638)$ (a) and  
$(T_\text{eff}=0.01, \phi=0.582)$ (b).
With increasing $\tau_p$ the height of all first peaks of $S(q)$ increases
at $T_\text{eff}=1.0$ while it decreases at $T_\text{eff}=0.01$. 
The qualitative evolution of the pair structure factor directly correlates with the 
evolution of the apparent glass transition lines in Fig.~\ref{fig:phasediagram}.}
\end{figure}

The contrast between the evolution of the density correlations with increasing
departure from equilibrium at high and low temperatures is easier
to observe in the Fourier domain. We show in Fig.~\ref{fig:sq} 
the partial static structure factor for the $A$ particles,
\begin{equation}
S(q) = \frac{1}{N_A}
\left< \sum_{i=1}^{N_A}\sum_{j=1}^{N_A} e^{-i\bm{q}\cdot(\bm{r}_i-\bm{r}_j)}\right>,
\end{equation} 
where both summations are over the $A$ particles. 

We observe in Fig.~\ref{fig:sq} that the amplitude of 
all the peaks of the static structure factor 
increases with increasing the persistence time for the higher
effective temperature, $T_\text{eff}=1.0$. In contrast, the amplitude of the 
peaks decreases with increasing the persistence time for the lower
effective temperature, $T_\text{eff}=0.01$. 
Therefore, the evolution of the peak height of the structure 
factor correlates directly with the evolution of the fitted glass transition 
volume fraction, $\phi_c$. We also find that the oscillations of the structure factor decay slower with 
increasing $\tau_p$, which directly reflects the sharpening and growth of the 
first peak of the pair correlation function discussed in Sec.~\ref{sec:gr} above.

The correlation identified between glassy dynamics and structure factor implies that to 
understand the gross features of nonequilibrium glass transitions in active systems, 
one must first understand how increasing departure from equilibrium affects the static structure 
of the active fluid at the level of two-point quantities. This first step is crucial, as some 
theories for the nonequilibrium glass transitions use the nonequilibrium pair correlation $S(q)$ 
as input for a dynamical theory of the driven glassy dynamics~\cite{Szamel2015,Szamel2016}.
This initial step is usually not emphasized in equilibrium studies~\cite{Berthier2011}, as there 
exist very accurate theories predicting the equilibrium $S(q)$ from the sole knowledge of the 
pair interaction~\cite{simple}.

\subsection{Physical interpretation: potential of mean force $u(r)$}

\label{sec:pmf}

There have been a few attempts to develop a theory for the static structure
of active fluids. One way to develop such a theory is to start with an 
approximate mapping of the nonequilibrium active fluid onto an effective
equilibrium fluid \cite{Farage2015,Maggi2015,Wittmann2017a,Wittmann2017b}. 
Having an effective equilibrium
model one could then use the well established framework of the equilibrium
liquid state theory to calculate the pair correlation function and the 
static structure factor. Farage \textit{et al.} \cite{Farage2015} showed that 
such a procedure results in very accurate predictions for the pair correlation
function, at least far from the glassy regime.
However, Rein and Speck \cite{Rein2016} found
that the structure of the simulated nonequilibrium active fluid can be quite
different from that of the simulated effective equilibrium fluid. 
Obviously, more work is needed in this area~\cite{recent-speck}. In particular, we note that
no comparison between functions characterizing density correlations obtained
from simulations and the same functions calculated using liquid state theory 
was performed in the glassy regime that is of interest for the present work.

A different way to evaluate the static structure of an active fluid is to
use the very recently proposed 
`integration through transients' approach~\cite{Liluashvili2017}
which was originally developed to describe sheared colloidal suspensions 
\cite{Fuchs2009}. So far, this approach was only used to analyze transient
glassy dynamics (\textit{i.e.} the dynamics after active forces have been turned on) but in 
principle it could be used to calculate equal-time
steady-state properties. 

In fact, to get some insight into the evolution of the density correlations
upon increasing the persistence time it would be nice to have an effective,
persistence time-dependent potential. In other words, we would like to 
investigate a potential that, if used in an equilibrium simulation, would
result in the nonequilibrium pair correlation functions shown in
Fig.~\ref{fig:gr}. Finding an effective potential resulting in a given
pair correlation function is a separate project. Moreover, 
it is not even obvious that such a potential could always be found
for our nonequilibrium pair correlation functions. 

\begin{figure}
\includegraphics[width=8.5cm]{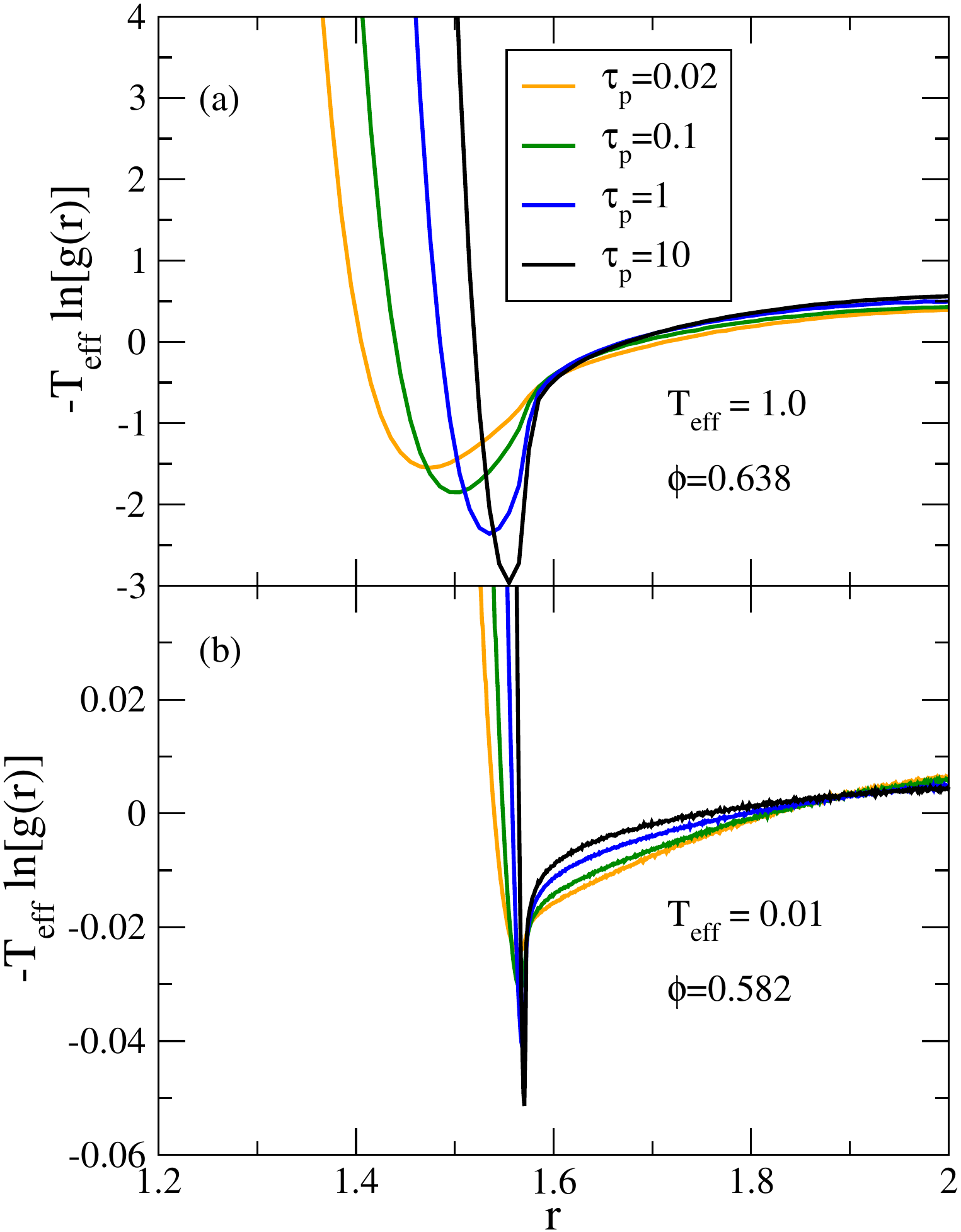}
\caption{\label{fig:pmf}  
Evolution of the potential of mean force  with the 
persistence time at $(T_\text{eff}=1.0, \phi=0.638)$ (a) and $(T_\text{eff}=0.01, \phi=0.582)$ (b). 
For $T_{\rm eff} = 1.0$, particle appear larger for larger $\tau_p$ (which makes the system more glassy), 
while they appear 
more sticky at $T_{\rm eff}=0.01$ (which fluidifies the system).}
\end{figure}

Instead, here we look at the evolution of the generalized potential of mean force, 
\begin{equation}
u(r) = - T_\text{eff} \ln [g(r)],
\end{equation} 
on the persistence time to obtain some physical intuition about the effective interaction experienced by the 
particles. Recall that $u(r)$ only represents a gross approximation to the real pair interaction between the 
particles, even at thermal equilibrium.  
We present numerical results for $u(r)$ in Fig.~\ref{fig:pmf}.
At both high and low temperatures, the evolution of $u(r)$ signals the emergence of short-range adhesive 
forces, manifested through the developement of a narrow negative well near the interparticle distance.   

However, at $T_\text{eff}=1.0$ the potential of mean force reveals a striking feature, that is not observed 
at $T_\text{eff}=0.01$. We observe that the repulsive interaction becomes much steeper with increasing the 
persistence time, and the spatial range of the repulsion extends to larger distances.  The particles, therefore, 
effectively become stiffer and,
perhaps more dramatically, they also appear larger with increasing persistence time. It is reasonable to 
assume that this apparent `expansion' of the particles directly accounts for the observed shift of the glass 
transition towards smaller volume fractions with increasing $\tau_p$ 
for $T_{\rm eff} = 1.0$. In other words, glassy dynamics is promoted in this regime because self-propulsion 
makes the active suspension effectively more crowded than its equilibrium counterpart.  
The evolution of the local structure for $T=1.0$ also explains the drastic decrease in the height of the 
intermediate-time plateau in the mean-squared displacements shown in Fig.~\ref{fig:msd}(a). 

For $T_\text{eff}=0.01$, an apparent expansion of the particles is barely visible. This is physically reasonable, 
as the equilibrium system is already very close to the hard sphere interaction, and it is difficult to 
make this hard-core interaction steeper. In this regime, the main effect of increasing the persistence 
time is the emergence of a short-range effective adhesion between the self-propelled particles. The emergence of 
adhesion through self-propulsion was noted before in the fluid regime~\cite{Ginot2015,Levis2014}, and 
we find here that it is quite pronounced for dense assemblies of self-propelled hard spheres as well, 
as surmised in Ref.~\cite{Berthier2014}. In that case, it is not surprising that
the glass transition is shifted to larger $\phi$, as this behavior 
is well-known from equilibrium studies of adhesive spheres \cite{pham}. 
Therefore, the physical reason that self-propelled hard spheres undergo a delayed glass transition 
is not that they are `driven' out of equilibrium by active forces, it is much more subtle. Instead, 
active forces induce an effective adhesion between the particles, which modifies the local structure of 
the fluid towards the one of equilibrium sticky spheres. It is this effective stickiness which eventually 
fluidifies the dense fluid.  
This interpretation also accounts for the subdiffusive plateau dynamics reported in Fig.~\ref{fig:msd}(b), 
which stems physically from the existence of two length scales controlling the intermediate time dynamics, 
the cage size and the adhesion range. 

Finally, we note that the above described changes in the potential of mean
force and the pair correlation function are not well captured by the 
approximate mappings of active systems onto effective equilibrium systems 
described in the literature.  
In particular, these approximations fail badly for low 
$T_\text{eff}$ as they fail to predict effective adhesion between the particles. 
They can, however, very qualitatively reproduce the effective `inflation' of the particles for soft interactions, 
and the qualitative change in the steepness. 
These results shows that deriving more accurate effective interactions for nonequilibrium active particles is an important research goal.    

\section{Discussion}\label{sec:disc}

Systems of interacting self-propelled particles undergo a glass transition which has 
many of the features associated with glass transitions in equilibrium systems. 
The transition is characterized by a dramatic 
slowing down upon a small change in control parameters, either temperature, 
effective temperature, or density. When glassy dynamics set in, the particles become 
confined in cages of neighboring particles. Consequently, a plateau develops in the 
mean-squared displacement and the intermediate scattering function. Importantly, 
profound changes in the dynamics are accompanied by 
very small changes in the static structure of the particles as observed in 
two-body correlation functions. 

There are, however, differences in the specifics of the transition that become apparent
as one increases the departure from equilibrium, which for our system is achieved
by increasing the persistence time $\tau_p$. For larger $\tau_p$ there is a decrease 
in the short-time ballistic dynamics due to velocity correlations. The effect of these 
velocity correlations on the short-time dynamics is correctly 
captured through a mode-coupling-like theory which we developed 
previously \cite{Szamel2015,Szamel2016}. 

Furthermore, increasing $\tau_p$ results in particles with smaller cages, which
is easily seen in the lower plateau height of the mean-squared displacements, 
and a larger effective radius, which is easily observed in an increase in the peak position of the 
pair-correlation function $g(r)$ at the same packing fraction. While the system continues to flow 
with increasing $\tau_p$, there are differences 
in the structure. However, at fixed $\tau_p$ these structural changes are small
and a glass transition still exists. The persistence time dependence of the
glass transition can be inferred from the decay of the oscillations of 
$g(r)$ at a fixed volume fraction. A faster decay indicates a larger
glass transition packing fraction. This feature can be observed more easily
in the evolution of the peak height of the static structure factor $S(q)$ with
increasing $\tau_p$. A smaller
height of $S(q)$ at fixed packing fraction leads to a larger glass
transition packing fraction. 

We can gain some insight into the effect of activity by examining
the potential of mean force $u(r) = -T_\text{eff} \ln[g(r)]$. 
While it is important to emphasize that, strictly speaking, 
$u(r)$ only has meaning for equilibrium systems, it does allow us 
to gain some insight into how to think about the effect of activity on
the behavior of systems. The potential of mean force is reminiscent 
of systems with short-range attractions, commonly called sticky spheres. 
With increasing $\tau_p$ the attractive range decreases, but the effective
particle size increases. The former effect is more pronounced at lower effective
temperatures where the width of the well of $u(r)$ is very small and the latter
effect seems to be the dominant feature at higher effective temperatures. 
This shows that predicting the effect of increasing departure from equilibrium is not trivial, and 
depends on the specifics of the studied model. 

Our conclusion is reminiscent of previous findings in equilibrium studies of the glass transition. 
When the pair structure evolves significantly by varying an external parameter, then the qualitative 
evolution of the location of the glass transition can be correctly inferred from the sole knowledge of 
pair density correlations. Good examples are the effect of increasing the adhesion strength in adhesive 
hard spheres~\cite{pham}, or increasing the density in ultra-soft colloidal particles~\cite{Moreno2012}. 
However, in order to quantitatively predict the location of the glass transition, it is well-established 
that two-point density functions are insufficient at equilibrium~\cite{gilles2}, and that higher-order 
correlations become relevant~\cite{Berthier2011}. This cautionary remark presumably also applies to active forces, 
and these more complicated correlations should also be studied in the present nonequilibrium context. 

Overall, our work demonstrates that nonequilibrium glass transitions appear robustly in dense active materials, 
and that much remains to be done at the theoretical level to derive predictive theories, even for very simple 
models of active particles. 

\section*{Acknowledgments} 

The research in Montpellier was supported by funding
from the European Research Council under the European
Union's Seventh Framework Programme (FP7/2007-2013) / ERC 
Grant agreement No 306845 and by a grant from
the Simons Foundation (\#454933, Ludovic Berthier).
E. F. and G. S. gratefully acknowledge the 
support of NSF Grant No.~CHE 1213401.

\end{document}